\documentclass[aip,apl,reprint]{revtex4-1}
\usepackage[T1]{fontenc}
\usepackage{times}
\usepackage{graphicx,color}
\usepackage{amsfonts,amsmath,amssymb,amsbsy}
\usepackage[colorlinks=true,linkcolor=blue,citecolor=blue,urlcolor=blue]{hyperref}
\def\emph#1{\textcolor{red}{#1}}

\begin{document}

\title{Signal detection based on the chaotic motion of an antiferromagnetic domain wall}

\author{Laichuan Shen}
\thanks{These authors contributed equally to this work.}
\affiliation{School of Science and Engineering, The Chinese University of Hong Kong, Shenzhen, Guangdong 518172, China}

\author{Jing Xia}
\thanks{These authors contributed equally to this work.}
\affiliation{College of Physics and Electronic Engineering, Sichuan Normal University, Chengdu 610068, China}

\author{Motohiko Ezawa}
\affiliation{Department of Applied Physics, The University of Tokyo, 7-3-1 Hongo, Tokyo 113-8656, Japan}

\author{Oleg A. Tretiakov}
\affiliation{School of Physics, The University of New South Wales, Sydney 2052, Australia}

\author{\\ Guoping Zhao}
\email[Email:~]{zhaogp@uestc.edu.cn}
\affiliation{College of Physics and Electronic Engineering, Sichuan Normal University, Chengdu 610068, China}

\author{Yan Zhou}
\email[Email:~]{zhouyan@cuhk.edu.cn}
\affiliation{School of Science and Engineering, The Chinese University of Hong Kong, Shenzhen, Guangdong 518172, China}

\begin{abstract}
The antiferromagnetic domain wall dynamics is currently a hot topic in mesoscopic magnetic systems. In this work, it is found that, based on the Thiele approach, the motion of an antiferromagnetic domain wall is described by the Duffing equation. Numerical simulations demonstrate that the antiferromagnetic domain wall can be used as a Duffing oscillator, and the transition between the periodic and chaotic motion can be used to detect the periodic signal in the presence of the white noise. Furthermore, we calculate the bifurcation diagram and Lyapunov exponents to study the chaotic behavior of an antiferromagnetic domain wall. The numerical simulations are in good agreement with the analytical solutions. Our results may be useful for building spintronic detection devices based on antiferromagnetic domain walls.
\end{abstract}

\date{\today}

\maketitle


%
Antiferromagnetic (AFM) materials are ordered spin systems, which are promising for building advanced spintronic devices due to their ultrafast spin dynamics and zero stray fields.~\cite{Baltz_RMP2018,Jungwirth_NNANO2016,Smejkal_NATP2018,Gomonay_NATP2018,Fukami_JAP2020}
The AFM spin textures, including AFM domain walls and skyrmions, can be controlled by various methods, such as by using spin currents,~\cite{Hals_PRL2011,Shiino_PRL2016,Velkov_NJP2016,Zhang_SREP2016A,Barker_PRL2016,Zhang_NATCOM2016,Yang_PRL2018,Jin_APL2016} magnetic fields,~\cite{Khoshlahni_PRB2019,Yuan_PRB2018,Gomonay_APL2016,Dasgupta_PRB2017} magnetic anisotropy gradients,~\cite{Shen_PRB2018,Yamada_APE2018} temperature gradients,~\cite{Khoshlahni_PRB2019,Chen_PRB2019,Selzer_PRL2016} and spin waves.~\cite{Qaiumzadeh_PRB2018,Tveten_PRL2014,Daniels_PRB2019}
In particular, the AFM domain walls located in the transition regions between AFM domains have no Walker breakdown due to the existence of the strong AFM exchange interaction, and their velocity can reach a few kilometers per second.~\cite{Shiino_PRL2016,Gomonay_PRL2016}
Recently, such ultra-fast motion of domain walls has been experimentally demonstrated in the ferrimagnetic $\text{Gd}_{3}\text{Fe}_{5}\text{O}_{12}$ film (it has a similar spin structure to antiferromagnet).~\cite{Zhou_arxiv2019}

For the AFM system, its dynamics are governed by two coupled Landau-Lifshitz-Gilbert (LLG) equations.~\cite{Gilbert_IEEE2004}
Based on such two first-order equations with respect to time, one can obtain a second-order equation for the AFM order parameter (i.e., the N{\'e}el vector).~\cite{Baltz_RMP2018}
Therefore, the AFM texture will acquire an effective mass and the equation of motion should be similar to Newton's kinetic equation.
As reported in a recent work,~\cite{Shen_APL2019} the motion of an AFM skyrmion in the nanodisk obeys the inertial dynamics, and its oscillation frequency may reach tens of GHz.

On the other hand, the LLG equation is nonlinear, which could lead to complex or even chaotic dynamic behaviors of the system.~\cite{Moon_SciRep2014,Yang_PRL2007,Devolder_PRL2019,Matsumoto_PRApplied2019,Petit-Watelot_NatPhys2012,Ohuno_JAP1997,Sukiennicki_JMMM1994,Matsushita_JPCJ2012}
For a chaotic system, its motion is sensitive to the initial conditions and cannot be predicted over a long time.
The chaotic systems play an important role in the applications.~\cite{Fukushima_APE2014,Ditto_Chaos2015,Wang_IEEE1999,Wu_MSSP2017}
For instance, considering a Duffing chaotic system, the transition from chaotic motion to periodic motion can be used to detect the periodic signal in the noisy environment.~\cite{Wang_IEEE1999}
The periodic signal detection in the noisy environment is widely applied to various fields, including secure communication, radar information detection, condition monitoring and fault diagnosis.~\cite{Wang_Kybernetes2009}
Although fast Fourier transform has the ability to extract the weak periodic signal from the noisy environment, the frequency of the to-be-detected signal cannot be determined accurately, while the chaotic oscillator can be used to determine the frequency accurately.~\cite{Wang_Kybernetes2009}
Interestingly, the motion of an AFM texture induced by alternating currents obeys the well-known Duffing equation, which describes the oscillation of an object with mass, as reported in Ref.~\onlinecite{Shen_Arx2019}.
Therefore, the AFM texture, such as the AFM domain wall, can be treated as a Duffing oscillator, which can be used in the signal detection.
However, the study of periodic signal detection based on the AFM domain wall is still lacking. 

In this work, we propose to use the motion of an AFM domain wall to detect the periodic signal in the noisy environment.
Our theoretical results show that the motion of an AFM domain wall can be described by the Duffing equation, and there is a transition between chaotic and periodic motion.
Based on such a transition, we propose a method to detect the frequency, phase and amplitude of the periodic signal.
Our numerical simulations prove the feasibility of using the AFM domain wall to detect the signal.

\begin{figure}[t]
\centerline{\includegraphics[width=0.48\textwidth]{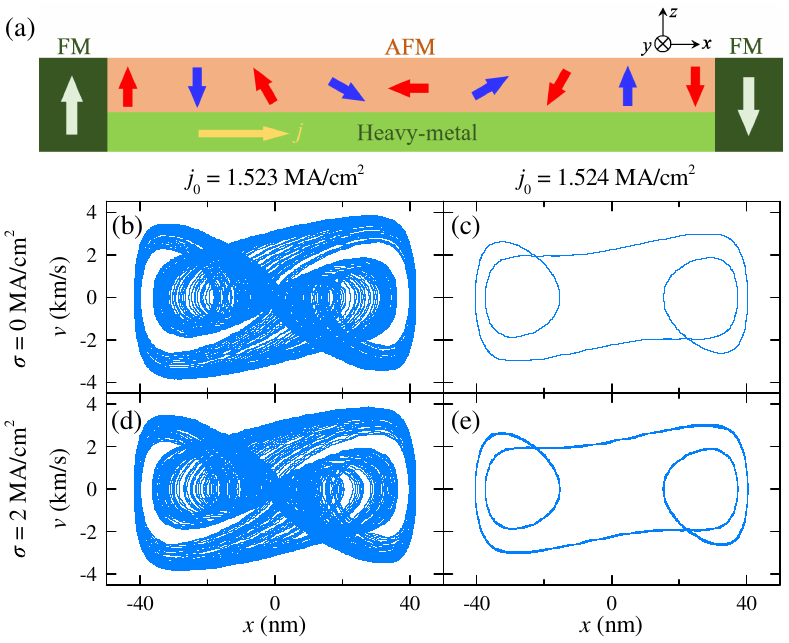}}
\caption{(a) The sketch of our calculation model. The heavy-metal layer is employed to generate the spin current. Such a spin current will apply spin torques to drive the AFM domain wall. In order to form the AFM domain wall and avoid the annihilation of the fast-moving domain wall at the AFM edge, two hard FM materials are considered. (b)-(e) Simulated trajectories of the AFM domain wall on the position ($x$)-velocity ($v$) plane, where the alternating current [$j = j_{0}$sin($2\pi ft$) with frequency $f = 7$ GHz and amplitude $j_{0} = 1.523$ and $1.524$ MA/cm$^{2}$] is applied, and the white noise with standard deviation $\sigma = 0$ and $2$ MA/cm$^{2}$ is added at $t$ = 2 ns. In our simulations, the following parameters are adopted,~\cite{Barker_PRL2016} $A=6.59$ pJ/m, $K=0.116$ MJ/m$^{3}$, $D=0.6$ mJ/m$^{2}$, $M_\text{S}=376$ kA/m, $\lambda=75.433$ MJ/m$^{3}$, $L=15.765$ mJ/m$^{2}$, and $\gamma=2.211 \times 10^{5}$ m/(A s). The damping $\alpha=0.0006$, which is a realistic value in antiferromagnets.~\cite{Moriyama_PRMat2019} Here, the change of magnetic moments only occurs in the $x$ direction, so that the mesh size of $1 \times 20 \times 1$ nm$^{3}$ can be used to discretize the AFM film with the size $100 \times 20 \times 1$ nm$^{3}$.}
\label{FIG1}
\end{figure}

%
We focus on the motion of the domain wall in the AFM layer, and the model is depicted in the Fig.~\ref{FIG1}(a).
The heavy-metal layer is employed in order to drive the AFM domain wall via spin-orbit torques (for the spin-transfer torque, it should also be applicable).~\cite{Shiino_PRL2016,Velkov_NJP2016}
In addition, two hard ferromagnets are considered for the following purposes.
The first purpose is to form the AFM domain wall by using the exchange coupling at the ferromagnetic (FM)/antiferromagnetic (AFM) interface,~\cite{Morales_PRL2015,Park_NatMater2011,Scholl_PRL2004,Nogues_JMMM1999,Lang_SciRep2018}
and the second purpose is to avoid the annihilation of the fast-moving domain wall at the AFM edge (see supplemental material).

Assuming that the directions of magnetic moments in the two-sublattice AFM film (with sublattice magnetization $\boldsymbol{M}_{\text{1}}$ and $\boldsymbol{M}_{\text{2}}$) vary along the $x$ axis only,
the AFM energy $E$ can be written as~\cite{Tveten_PRB2016,Shiino_PRL2016} $E = \int{\mathcal{F}} \,dV$,
where $\mathcal{F}=\frac{\lambda}{2}\boldsymbol{m}^{\text{2}}+\frac{A}{2}(\partial_{x}\boldsymbol{n})^{\text{2}}+L\boldsymbol{m}\cdot\partial_{x}\boldsymbol{n}-\frac{K}{2}(\boldsymbol{n}\cdot\boldsymbol{n_{e}})^{2}+\frac{D}{2}\boldsymbol{e_{y}}\cdot(\boldsymbol{n}\times\partial_{x}\boldsymbol{n})$
with the homogeneous exchange constant $\lambda$, inhomogeneous exchange constant $A$, parity-breaking constant $L$~\cite{Tveten_PRB2016,Shiino_PRL2016,Qaiumzadeh_PRB2018}, magnetic anisotropy constant $K$ and Dzyaloshinskii-Moriya interaction (DMI) constant $D$~\cite{Dzyaloshinsky_JPCS1958,Moriya_PR1960,Rohart_PRB2013}.
$\boldsymbol{n_{e}}=\boldsymbol{e_{z}}$ stands for the direction of the anisotropy axis.
$\boldsymbol{n}=(\boldsymbol{m}_{\text{1}}-\boldsymbol{m}_{\text{2}})/2$ and $\boldsymbol{m}=(\boldsymbol{m}_{\text{1}}+\boldsymbol{m}_{\text{2}})/2$ are the staggered magnetization (or N{\'e}el vector) and the total magnetization,
where $\boldsymbol{m}_{i}$ ($=\boldsymbol{M}_{i}/M_\text{S}$ with the saturation magnetization $M_\text{S}$) is the reduced magnetization. 
For most realistic cases where the AFM exchange interaction is significantly strong, $\boldsymbol{m}^{\text{2}} \ll \boldsymbol{n}^{\text{2}} \sim {1}$.~\cite{Dasgupta_PRB2017,Zarzuela_PRB2018,Keesman_PRB2016}
Due to the presence of the exchange coupling at the FM/AFM interface, the interface energy should be introduced, so that $E \to E-J_{\text{int}}\int{\boldsymbol{n_{AF}}\cdot\boldsymbol{m_{F}}} \,dS$,
where $J_{\text{int}}$ is the interfacial exchange coupling constant, and $\boldsymbol{n_{AF}}$ and $\boldsymbol{m_{F}}$ are the AFM and FM magnetic moments at the interface respectively.
The variational derivatives of the AFM energy can give the static profiles of the AFM domain wall, as shown in supplemental material.

Taking the spin-orbit torques (SOTs) into account, the equations of motion are described as~\cite{Shiino_PRL2016,Velkov_NJP2016,Gomonay_PRB2010}
%
\begin{align}
\boldsymbol{\dot{n}}&=(\gamma\boldsymbol{f}_{m}-\alpha\boldsymbol{\dot{m}})\times\boldsymbol{n}+\boldsymbol{T}_{n,\text{SOT}},\tag{1a}\label{eq:1a}\\
\boldsymbol{\dot{m}}&=(\gamma\boldsymbol{f}_{n}-\alpha\boldsymbol{\dot{n}})\times\boldsymbol{n}+\boldsymbol{T}_{nl}+\boldsymbol{T}_{m,\text{SOT}},\tag{1b}\label{eq:1b}
\end{align}
%
where $\gamma$ and $\alpha$ are the gyromagnetic ratio and the damping constant respectively, and $\boldsymbol{T}_{nl}=(\gamma\boldsymbol{f}_{m}-\alpha\boldsymbol{\dot{m}})\times\boldsymbol{m}$ is the higher-order nonlinear term~\cite{Hals_PRL2011}. 
$\boldsymbol{T}_{n,\text{SOT}}=\gamma H_{j}\boldsymbol{m}\times\boldsymbol{p}\times\boldsymbol{n}$ and $\boldsymbol{T}_{m,\text{SOT}}=\gamma H_{j}\boldsymbol{n}\times\boldsymbol{p}\times\boldsymbol{n}$ are damping-like spin-orbit torques,
where $\boldsymbol{p}$ is the polarization vector and $H_{j}$ relates to the applied current density $j$, defined as $H_{j}=j\hbar P/(2\mu_{0}eM_\text{S}t_{z})$ with reduced Planck constant $\hbar$, spin-Hall angle $P$, vacuum permeability constant $\mu_{0}$, elementary charge $e$, and layer thickness $t_{z}$.
In this work, we focus on the study of detecting the periodic current signal in the noisy environment, and the to-be-detected signal and white noise are added to the applied current $j$.
$\boldsymbol{f}_{n}=-\delta E/\mu_{\text{0}}M_\text{S}\delta\boldsymbol{n}$ and $\boldsymbol{f}_{m}=-\delta E/\mu_{\text{0}}M_\text{S}\delta\boldsymbol{m}$ are the effective fields.

%
Using Eqs.~(\ref{eq:1a}) and~(\ref{eq:1b}), we simulate the motion of a domain wall in the AFM film (details of the simulations are given in the supplemental material).
To track the AFM domain wall, $x = \int{x(1-n_z^{2})}dx/\int{(1-n_z^{2})}dx$ is used.
Figures~\ref{FIG1}(b)-(e) show that the AFM domain wall exhibits different motion behavior for $j_{0} = 1.523$ and $1.524$ MA/cm$^{2}$,
where the alternating current [$j = j_{0}$sin($2\pi ft$) with frequency $f = 7$ GHz and amplitude $j_{0}$] is used as the driving source.
For the case of $j_{0} = 1.523$ MA/cm$^{2}$, the motion of the AFM domain wall is chaotic (for the chaotic motion, the time evolution of position of the AFM domain wall has been plotted in the supplemental material), while for $j_{0} = 1.524$ MA/cm$^{2}$, it is periodic.
On the other hand, as shown in Fig.~\ref{FIG1}, even if there is the white noise with standard deviation $\sigma = 2$ MA/cm$^{2}$,
the transition between chaotic and periodic motion does not occur,
so that the AFM system studied here has the immune ability to the noise.

%
In order to analyze the above motion behavior of the AFM domain wall, we will derive the steady motion equation.
By using Thiele (or collective coordinate) approach~\cite{Thiele_PRL1973,Tveten_PRL2013,Tretiakov_PRL2008,Clarke_PRB2008,Zarzuela_PRB2018}, the steady motion equation for an AFM domain wall is obtained from Eqs.~(\ref{eq:1a}) and~(\ref{eq:1b}), written as (see supplemental material for details)
\begin{equation}
M_{\text{eff}}\ddot{x}+\alpha^{*}\dot{x}=F_{\text{SOT}}+F_{b},\tag{2}
\label{eq:2}
\end{equation}
where $x$ denotes the position of the AFM domain wall, and $M_\text{eff}$ is the effective AFM domain wall mass, which is defined as $\mu_{0}^{2}M_\text{S}^{2}t_{y}t_{z}d/\gamma^{2}\lambda$ with width $t_{y}$ of the AFM layer.
The second term on the left side of Eq.~(\ref{eq:2}) relates to the dissipative force, where $\alpha^{*}=\alpha\mu_{0}M_\text{S}t_{y}t_{z}d/\gamma$ and $d = \int{dx}(\partial_{x}\boldsymbol{n}\cdot\partial_{x}\boldsymbol{n})$.
$F_{\text{SOT}}$ is the force induced by SOTs, $F_{\text{SOT}}=-\mu_{0}H_{j}M_\text{S}t_{y}t_{z}\int{dx}[(\boldsymbol{n}\times\boldsymbol{p})\cdot\partial_{x}\boldsymbol{n}]$.
For the alternating current $j = j_{0}$sin($2\pi ft$), $F_{\text{SOT}} = F_{\text{SOT},0}\text{sin}(2\pi ft)$ with $F_{\text{SOT},0} \approx \pi \mu_{\text{0}}H_{j}M_\text{S}t_{y}t_{z}$.
$F_{b}$ stands for the boundary-induced force, which can be described by the polynomial $F_{b} \approx -b_{1}x-b_{2}x^{3}-b_{3}x^{5}-b_{4}x^{7}$ for the AFM film with length of $100$ nm studied here (see supplemental material for details on these values of $b_{1}, b_{2}, b_{3}$ and $b_{4}$).
Note that for the chaotic behavior, the presence of the $x^{3}$ term is enough, while in order to match the numerical results, the $x^{5}$ and $x^{7}$ terms are introduced.
Since $F_{b}$ contains the nonlinear terms, Eq.~(\ref{eq:2}) is called the (modified) Duffing equation~\cite{Novak_PRA1982,Moon_SciRep2014},
which describes the oscillation of an object with mass under the action of nonlinear restoring forces.
In this work, the thermal fluctuations~\cite{Brown_PR1963,Bessarab_PRB2019} are not considered.
If there is the thermal fluctuation, the random thermal force should be added in Eq.~(\ref{eq:2}),~\cite{Lin_PRB2013,Brown_PRB2019} in order to analyze the effect of thermal fluctuation on the motion of the AFM domain wall.  

\begin{figure}[t]
\centerline{\includegraphics[width=0.48\textwidth]{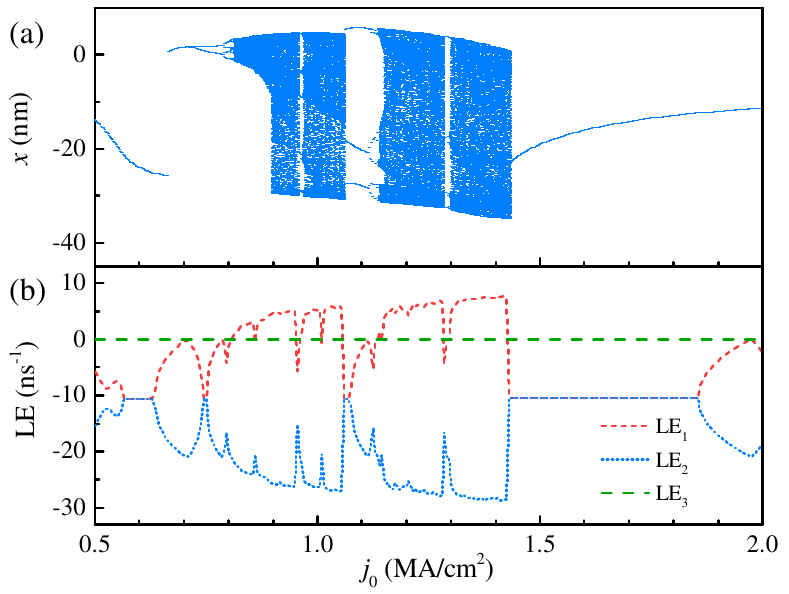}}
\caption{(a) Calculated bifurcation diagram, where the damping $\alpha$ is 0.0006, and the alternating current with frequency $f = 7$ GHz and amplitude $j_{0} = 0.5 \sim 2.0$ MA/cm$^{2}$ is adopted. (b) Lyapunov exponents (LEs) as functions of the amplitude $j_{0}$.}
\label{FIG2}
\end{figure}
%
Based on Eq.~(\ref{eq:2}), we calculate the bifurcation diagram by using stroboscopic sampling for every driving period, as shown in Fig.~\ref{FIG2}(a).
Figure~\ref{FIG2}(a) shows a jump at $j_{0} = 0.66$ MA/cm$^{2}$, which is a common phenomenon in nonlinear systems with multistability.~\cite{Pivano_PRB2016}
As the amplitude $j_{0}$ of the alternating current increases, the period-doubling phenomenon occurs, and then the motion of the AFM domain wall shows chaotic behavior.
When $j_{0}$ increases to the critical value $j_{c}$ of $\sim 1.434$ MA/cm$^{2}$, the chaotic motion will become the periodic motion.
Such a transition from chaotic to periodic motion is reproduced by the numerical simulations, as shown in Fig.~\ref{FIG1},
where the critical value $j_{c}$ ($\sim 1.523$ MA/cm$^{2}$) obtained from the numerical simulations is close to that ($\sim 1.434$ MA/cm$^{2}$) given by Eq.~(\ref{eq:2}).
In addition, the critical current $j_{c}$ increases with the frequency (see supplemental material for details).

On the other hand, the Lyapunov exponents (LEs) are usually used to judge whether there is chaos, given as~\cite{Yang_PRL2007,Souza-Machado_AmJP1990,Shen_Arx2019} 
\begin{equation}
\text{LE}_{i}=\lim\limits_{t \to \infty} \frac{1} {t} \text{ln} \frac{\left \| \delta x^{i}_{t} \right \|} {\left \| \delta x^{i}_{0} \right \|}. \tag{3}
\label{eq:3}
\end{equation}
For the nonlinear system studied here, it is a three-dimensional autonomous system, so that $i = 1, 2, 3$.~\cite{Shen_Arx2019}
$\left \| \delta x^{i}_{0} \right \|$ is the distance between two close trajectories at initial time,
and $\left \| \delta x^{i}_{t} \right \|$ is the distance between the trajectories at time $t$.
If the largest LE is positive, the attractor for the system is strange (or chaotic), two close trajectories will be separated and a small initial error increases rapidly,
resulting in that the motion is sensitive to the initial condition and shows chaotic behavior.
Figure~\ref{FIG2}(b) shows our calculated LE, which is consistent with the bifurcation diagram.
The sum of LEs equals to $-\alpha^{*}/M_{\text{eff}}=-\alpha\lambda\gamma/\mu_{0} M_{\text{S}}$ (it is $-21.18$ ns$^{-1}$ for $\alpha = 0.0006$),
indicating that a small damping $\alpha$ can lead to the chaotic behavior (the effect of the damping on the bifurcation diagram and LEs has been shown in the supplemental material).~\cite{Shen_Arx2019}
For the case of the presence of the white noise, as reported in Ref.~\onlinecite{Wu_MSSP2017}, the influence of the noise on LEs maintains the characteristics of zero, so that the Duffing systems have strong noise immunity.

For an AFM domain wall, its equation of motion can be described as the Duffing equation, i.e., Eq.~(\ref{eq:2}).
Therefore, the AFM domain wall can be used as a Duffing oscillator,
and using the transition from periodic to chaotic motion (or from chaotic to periodic motion) could be employed to detect the periodic signal in the noisy environment.
Next, the method to detect the frequency, phase and amplitude of the periodic signal is introduced in detail.

We assume that the motion of the AFM domain wall is chaotic only under the action of the reference signal $j = j_{0}$sin($2\pi ft + \varphi$).
When the periodic signal $s = s_{0}$sin($2\pi f_{s}t + \beta$) is added, the total signal $j_{t}$ equals to $j_{t} = j + s = j_{t,0}$sin($2\pi ft + \varphi + \epsilon$), where the amplitude $j_{t,0}$ is
\begin{equation}
j_{t,0}=\sqrt{j_{0}^{2}+s_{0}^{2}+2j_{0}s_{0}\text{cos}[2\pi(f_{s}-f)t+(\beta-\varphi)]}, \tag{4}
\label{eq:4}
\end{equation}
and the change in phase is denoted by $\epsilon$,
\begin{equation}
\text{tan}\epsilon=\frac {s_{0}\text{sin}[2\pi(f_{s}-f)t+(\beta-\varphi)]} {j_{0}+s_{0}\text{cos}[2\pi(f_{s}-f)t+(\beta-\varphi)]}. \tag{5}
\label{eq:5}
\end{equation}
Usually, $j_{0} \gg s_{0}$, resulting in $\epsilon \sim 0$, so that the change of phase can be safely disregarded.
It can be seen from Eq.~(\ref{eq:4}) that the periodic signal will affect the amplitude $j_{t,0}$.
If $j_{t,0}$ exceeds the critical value $j_{c}$, the transition from chaotic to periodic motion will occur.   
By using such a transition, we can obtain the frequency $f_{s}$, phase $\beta$ and amplitude $s_{0}$ of the periodic signal. 

%
\begin{figure*}[t]
\centering    
\centerline{\includegraphics[width=0.96\textwidth]{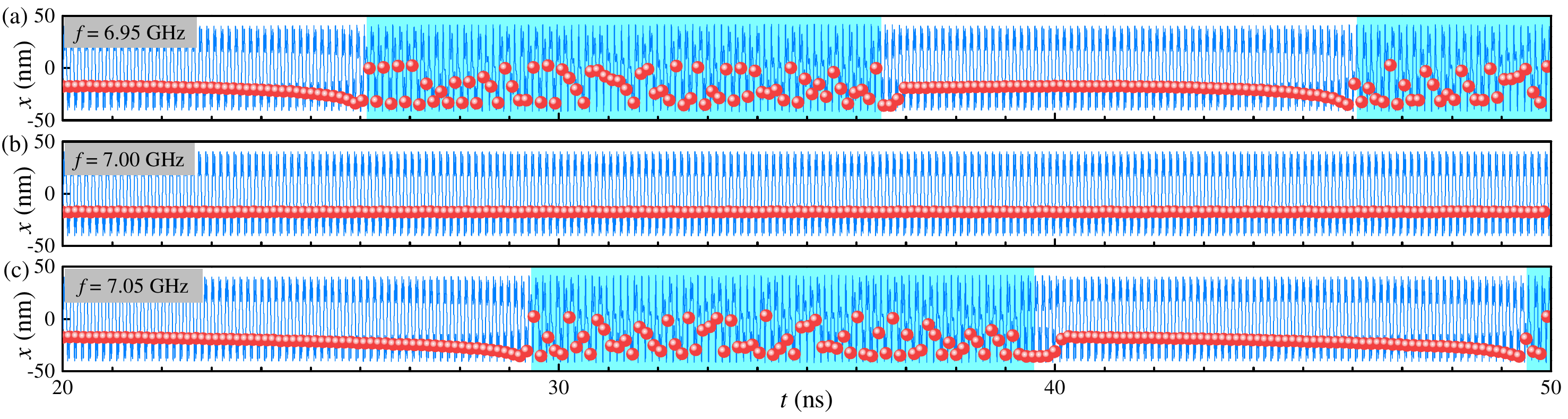}}
\caption{The time evolution of position for an AFM domain wall driven by alternating currents $j = j_{0}$sin($2\pi ft$) with different amplitudes and frequencies [(a) $f$ = 6.95 GHz and $j_{0}$ = 1.498 MA/cm$^{2}$; (b) $f$ = 7.00 GHz and $j_{0}$ = 1.523 MA/cm$^{2}$; (c) $f$ = 7.05 GHz and $j_{0}$ = 1.549 MA/cm$^{2}$]. In our simulations, the periodic current signal to be detected, i.e., $s = s_{0}$sin($2\pi f_{s}t + \beta$) with $s_{0} = 0.08$ MA/cm$^{2}$, $f_{s} = 7$ GHz and $\beta = 30^{\circ}$, is buried in the white noise with standard deviation $\sigma = 2$ MA/cm$^{2}$. The white noise and to-be-detected current signal are added at $t$ = 2 ns. The lines present the continuous evolution of the position $x$, while the symbols show the discrete position obtained by using stroboscopic sampling for every driving period. The color background indicates that the motion of the AFM domain wall is chaotic. }
\label{FIG3}
\end{figure*}

In order to detect the frequency $f_{s}$, it is necessary to construct an array of Duffing oscillators with different reference frequencies $f$.~\cite{Wang_IEEE1999}
If there is a frequency difference between $f_{s}$ and $f$, Eq.~(\ref{eq:4}) indicates that $j_{t,0}$ is periodically more than or less than $j_{c}$, where the period $T$ of the cycle is equal to,~\cite{Wang_IEEE1999}
\begin{equation}
T=\frac {1} {f_{s}-f}. \tag{6}
\label{eq:6}
\end{equation}
Thus, in the presence of the frequency difference ($f_{s} - f$), the intermittent chaos will take place.
To verify the above result, we simulate the motion of the AFM domain wall driven by alternating currents with different frequencies.
As shown in Figs.~\ref{FIG3}(a) and (c), the intermittent chaos is presented and the period $T$ is about 20 ns, as expected by Eq.~(\ref{eq:6}),
where the reference frequencies $f$ in Figs.~\ref{FIG3}(a) and (c) are 6.95 and 7.05 GHz respectively, and the frequency $f_{s}$ of the to-be-detected signal $s$ is set to 7 GHz.
For the case of $f = f_{s} = 7$ GHz, the intermittent chaos disappears [see Fig.~\ref{FIG3}(b)].

%
\begin{figure}[b]
\centerline{\includegraphics[width=0.48\textwidth]{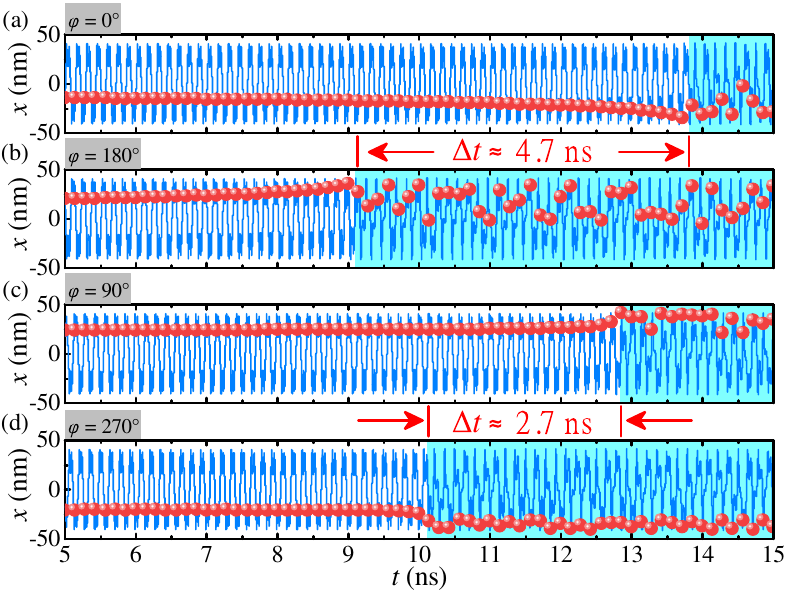}}
\caption{The time evolution of position for an AFM domain wall driven by alternating currents $j = j_{0}$sin($2\pi ft + \varphi$) with different phases [(a) $\varphi = 0^{\circ}$; (b) $\varphi = 180^{\circ}$; (c) $\varphi = 90^{\circ}$; (d) $\varphi = 270^{\circ}$;]. In our simulations, $f$ = 7 GHz, and $j_{0}$ changes linearly with the time $t$, i.e., $j_{0} = j_{0,0} - \frac{dj_{0}} {dt} \cdot t $, where $j_{0,0}$ = 1.8 MA/cm$^{2}$ and $\frac{dj_{0}} {dt} = 0.03$ MA/(cm$^{2}$ $\cdot$ ns). The periodic signal to be detected and the white noise are the same as those of Fig.~\ref{FIG3}.} 
\label{FIG4}
\end{figure}

We now study how to detect the phase $\beta$ and amplitude $s_{0}$ of the periodic signal.
For the case of $f = f_{s}$, using $j_{t,0} = j_{c}$ and combining Eq.~(\ref{eq:4}), it is found that when the amplitude $j_{0}$ of the reference signal arrives at the value of $j_{0,\varphi} = -s_{0}\text{cos}(\beta-\varphi)+\sqrt{j_{c}^{2}-s_{0}^{2}\text{sin}^{2}(\beta-\varphi)}$, the transition will occur.
Such a value, i.e., $j_{0,\varphi}$, shows that for different phases $\varphi$ of the reference signal $j$, different amplitudes $j_{0}$ of $j$ are required in
order to induce the occurrence of the transition.
For the phase $\varphi = 0^{\circ}$ and $180^{\circ}$, from the formula of $j_{0,\varphi}$, we get 
\begin{equation}
j_{0,\varphi = 180^{\circ}}-j_{0,\varphi = 0^{\circ}} = 2 s_{0}\text{cos}\beta. \tag{7}
\label{eq:7}
\end{equation}
For $\varphi = 90^{\circ}$ and $270^{\circ}$, the following equation is attained similarly,
\begin{equation}
j_{0,\varphi = 270^{\circ}}-j_{0,\varphi = 90^{\circ}} = 2 s_{0}\text{sin}\beta. \tag{8}
\label{eq:8}
\end{equation}
Thus, by scanning the amplitude $j_{0}$ of the reference signal $j$, one can get $j_{0,\varphi}$ for different $\varphi$ (if $j_{0} = j_{0,\varphi}$, the transition occurs),
and then, using Eqs.~(\ref{eq:7}) and~(\ref{eq:8}) gives the phase $\beta$ and amplitude $s_{0}$ of the periodic signal $s$.  
Figure~\ref{FIG4} shows the results of our numerical simulations, where $j_{0}$ changes linearly with the time $t$, i.e., $j_{0} = j_{0,0} - \frac{dj_{0}} {dt} \cdot t $ with $j_{0,0}$ = 1.8 MA/cm$^{2}$ and $\frac{dj_{0}} {dt} = 0.03$ MA/(cm$^{2}$ $\cdot$ ns).
For the case of Fig.~\ref{FIG4}, $j_{0,\varphi = 180^{\circ}}-j_{0,\varphi = 0^{\circ}} = \frac{dj_{0}} {dt} \cdot \Delta t = 0.141$ MA/cm$^{2}$ and $j_{0,\varphi = 270^{\circ}}-j_{0,\varphi = 90^{\circ}} = 0.081$ MA/cm$^{2}$. Note that there are small errors from the artificial choice of the boundary between chaotic and periodic motion. Substituting the above results into Eqs.~(\ref{eq:7}) and~(\ref{eq:8}), the phase $\beta = 29.9^{\circ}$ and amplitude $s_{0} = 0.081$ MA/cm$^{2}$ of the signal $s$ are calculated , which are consistent with the input values of $30^{\circ}$ and 0.08 MA/cm$^{2}$ in our simulations.


%
In summary, we propose to use the motion of an AFM domain wall to detect the periodic signal.
From the Thiele equation, Eq.~(\ref{eq:2}), we calculate the bifurcation diagram and Lyapunov exponents, and analyze the chaotic behavior of the AFM domain wall.
Moreover, our numerical simulations prove that using the transition between periodic and chaotic motion can be employed to detect the periodic signal in the presence of the white noise.
The results based on AFM domain walls can be extended to other types of AFM textures, such as skyrmion and bimeron,~\cite{Shen_Arx2019,Gobel_PRB2019,Leonov_PRB2017,Ezawa_PRB2011,Zhang_SREP2015,Kharkov_PRL2017,Kim_PRB2019,Woo_Nat2018,Yu_Nat2018,Zhou_NSR2018,Zhang_Arx2019,Nagaosa_NNANO2013,Fert_NATREVMAT2017,Kang_PIEEE2016} since their equations of motion are similar to Eq.~(\ref{eq:2}).
Although the topologically nontrivial skyrmions and bimerons can show rich dynamic phenomena (for example, the skyrmion Hall effect~\cite{Jiang_NatPhys2017,Litzius_NatPhys2017}), their formation and stability depend heavily on the parameters (generally, suitable DMI, magnetic anisotropy and exchange constants are required). Compared to skyrmions and bimerons, domain walls are easier to form and stabilize (domain wall can be stabilized even if there is no DMI).
Our results may provide guidelines for building spintronic detection devices based on the AFM textures.

See the supplemental material for details of the micromagnetic simulations and the analytical derivations for AFM domain walls.
The data that support the findings of this study are available from the corresponding authors upon reasonable request.


\textit{Acknowledgments.}
M.E. acknowledges the support by the Grants-in-Aid for Scientific Research from JSPS KAKENHI (Grant Nos. JP18H03676 and JP17K05490) and the support by CREST, JST (Grant Nos. JPMJCR16F1 and JPMJCR1874).
O.A.T. acknowledges the support by the Australian Research Council (Grant No. DP200101027), NCMAS 2020 grant, and the Cooperative Research Project Program at the Research Institute of Electrical Communication, Tohoku University.
G.Z. acknowledges the support by the National Natural Science Foundation of China (Grant Nos. 51771127, 51571126 and 51772004) and the Scientific Research Fund of Sichuan Provincial Education Department (Grant Nos. 18TD0010 and 16CZ0006).
Y.Z. acknowledges the support by the President's Fund of CUHKSZ, Longgang Key Laboratory of Applied Spintronics, National Natural Science Foundation of China (Grant Nos. 11974298 and 61961136006), Shenzhen Key Laboratory Project (Grant No. ZDSYS201603311644527), and Shenzhen Peacock Group Plan (Grant No. KQTD20180413181702403).



\end{document}